\documentclass[a4paper,twocolumn]{esapub} 

\usepackage{times}
\usepackage{natbib}
\usepackage{graphicx}

\title{Current status of Asteroseismology}

\author{Hans Kjeldsen}
\affil{Teoretisk Astrofysik Center, Aarhus Universitet, 8000 Aarhus C,
Denmark, hans@ifa.au.dk}
\author{Timothy R. Bedding}
\affil{School of Physics A28, University of Sydney, NSW 2006, Australia}
 
\begin{document}
 

\maketitle

\begin{abstract}
Oscillation frequencies are the most accurate properties one can measure
for a star, potentially allowing detailed tests of stellar models and
evolution theories.  We briefly review asteroseismology for two classes of
stars.  In $\delta$~Scuti variables, the main problems are the
identification of the observed modes and the theoretical treatment of
rotation.  In solar-like stars the main difficulty is the tiny amplitudes,
but credible detections are now being made.  These confirm that stars are
oscillating at the approximately the expected levels, but suggest that
amplitudes scale as $1/g$ rather than $L/M$.  We also stress the importance
of multi-site campaigns.  Several space missions will be launched over the
coming years, promising an exciting future for asteroseismology.
\end{abstract}

\section{Why is asteroseismology so far behind helioseismology?}

Asteroseismology involves an interplay between observations of stellar
oscillations and theoretical model calculations.  It can be done when a set
of oscillation frequencies is known for a given star and, at the same time,
a set of theoretical model frequencies can be calculated.  The motivation
for doing asteroseismology is that oscillation frequencies are the most
accurate properties one can measure for a star and we may therefore, at
least in principle, be able to perform a detailed test of stellar modelling
and evolution theories.
 
This potentially very promising tool has motivated a huge observational
effort with the aim of obtaining accurate oscillation frequencies.  At the
same time, a substantial amount of work has been put into improvements of
stellar modelling with the goal of being able to fit model frequencies to
the observed oscillation frequencies.
 
In helioseismology, as can be seen from the papers presented in these
proceedings, most current work concentrates on the details.  We are already
able to measure accurately many important properties of the solar interior.
Unlike asteroseismology, helioseismology is founded on an immense amount of
high-quality data supported by an equally detailed set of state-of-the-art
models.
 
It is impossible to imagine that asteroseismology will ever reach a level
similar to that in which we find helioseismology today, concerning analysis
techniques, data quality and the level of the results.  The reasons for
this lie in the main differences between the two subjects:
\begin{itemize}\itemsep=0pt  \vspace*{-2ex}

\item Asteroseismology works on distant stars whose basic properties will
always be less well determined than those of the Sun.  These include age,
composition, radius, mass, atmospheric properties and neutrino flux.  The
uncertainties will affect the quality of the calculated stellar models.
One example is age, which is known for the Sun from radioactive dating of
the solar system.  To illustrate the importance of this parameter for the
current solar model, consider the following simple question: What age would
we determine for a solar model -- based on the observed properties of the
Sun -- if we did not already know the age of the solar system?

\item The surfaces of stars are essentially unresolved, which limits
asteroseismology to modes of low degree.  Much of what we have learned
about the Sun is based on high degree modes.

\item The Sun is a relatively simple system.  This makes it very
interesting, since we may have a chance to understand it!  Many stars seem
to be much more complicated.  Hopefully we will find some important
physical properties that are not known at present, which may turn out to be
important for understanding the details of those stars.

\item There is only one Sun, while there are billions of other stars.  Many
people work on the details in helioseismology, but we can't expect several
billion astronomers to be working in asteroseismology!

\item In helioseismology we see some big networks and dedicated telescopes
(GONG, IRIS, BiSON), and we have many years of space research (SOHO,
SolarMAX, IPHIR).  In asteroseismology we find many smaller campaigns and
networks, but very few dedicated telescopes.  The space projects are just
beginning.  So far, there is a tremendous lack of high-quality data.

\item Asteroseismology works on fainter stars and so has lower sensitivity
than helioseismic observations.  One will therefore generally be restricted
to oscillations with relatively high amplitudes.

\end{itemize}
 
To explain why asteroseismology is so far behind helioseismology, we could
identify three important elements from the above list:
\begin{itemize}\itemsep=0pt  \vspace*{-2ex}
\item The imprecision of basic stellar properties.
\item The complication of the stellar physics.
\item The lack of high-quality oscillation data.
\end{itemize}
 
\section{A review of asteroseismology}
 
In this review we will discuss what has been learned and finish by
evaluating what we can hope to do in the next 10 years.
When trying to review such a huge scientific field, there is always the
risk of being remembered for the areas we did not discuss.  So let us begin
by saying what we do not intend to review.  The application to white dwarfs
is probably the biggest success of asteroseismology \citep[see, e.g.,][]{Vau97,Kaw98}, but we do not intend to mention any techniques and
results from this field.  The same is true for a number of classical
pulsating stars, such as $\beta$~Cephei stars and roAp stars, which will
not be discussed in this review.  On seismology of giants and subgiants we
refer to a paper by Guenther in these Proceedings.
 
We will discuss two areas of asteroseismology:
\begin{itemize}

\item Oscillations in $\delta$~Scuti stars, which are pulsating A and F
stars found on the main sequence or in the subgiant phase of their
evolution, inside the classical instability strip.  They have masses
between 1.5 and 2.5 solar masses.

\item Solar-like oscillations, which are conventionally defined as those
excited stochastically by turbulent convection \citep[e.g.,][]{HBChD99}.
These are expected in all stars on the cool side of the $\delta$~Scuti
instability strip, since it is these stars which have significant
convection in their outer regions.

\end{itemize}  
 
It is hoped that asteroseismology on these two classes of stars should
answer a number of central questions related to stellar structure and
evolution:
\begin{itemize}

\item Where, exactly, is the border between stars with and without a
convective core?  We believe it to be at a mass just slightly higher than
the solar (around 1.1\,$M_\odot$).

\item Stellar modelling suggest that stars on the main sequence can be
divided into non-evolved stars that contain fusion in the core of the star
and evolved subgiant stars that contain a fusion zone in a shell around the
core.  We expect to be able to locate the exact boarder between star having
core and shell hydrogen fusion.

\item In general we of course expect to test details of the models and also
be able to test a number of special features such as mixing, diffusion,
magnetic fields and rotation.

\item Finally, one may hope to be able to measure stellar ages.

\end{itemize}
 
\section{$\delta$~Scuti stars}

Asteroseismology of $\delta$~Scuti stars has been reviewed several times
\citep{Mat93,Daep93,Han2000} and there have also been workshops dedicated
to these objects \citep{B+MVienna2000}.  The number of detected and
accurately known frequencies is very large, but these are only useful if
one is able to calculate an equally accurate model frequency, which can
only be done if the mode has been identified.  Mode identification has been
one of the major obstacles to producing reliable seismic results on
$\delta$~Scuti stars.  A related problem is the fact that only a small
fraction of the possible oscillation modes seem to be excited to a
detectable level in a typical $\delta$~Scuti star.  Several techniques to
overcome the mode identification problem have been tried. One should in
this respect be careful about assumptions (in some cases even prejudice)
that are used to eliminate the mode identification problem.  However
probably only a fully open-minded attitude will move us towards a solution
to this serious problem.
 
\subsection{Fitting frequencies without knowing the identity of the modes}

A way through this has been to calculate a huge set of models without
assuming any identity for any of the observed modes, and then by a simple
$\chi^2$ calculation, locate the optimised solution.  This approach have
been used by \citet{PDH98} for the $\delta$~Scuti star XX Pyxidis.  Based
on frequencies for 13 oscillation modes in XX Pyx, they constructed 40\,000
sets of model frequencies based on stellar evolution calculations that
included rotation.  In the search for the best input model and the
optimised mode identification, \citeauthor{PDH98} were able to locate 8
local minima in $\chi^2$ space corresponding to 8 quite different
solutions.  Although they identified these as reasonable solutions, it is
also clear than none of the 40\,000 sets of models frequencies was able to
reproduce exactly the observed frequencies.  The reason is probably that
the physics in the models does not describe the real properties of the
star.  It therefore seems clear that mode identification is needed before
one can move forward in constructing seismic models of $\delta$~Scuti
stars.
 
\subsection{Mode identification in FG~Vir}

A new technique for mode identification was introduced by \citet{VKB98},
based on oscillation amplitudes measured via changes in the equivalent
widths of hydrogen and metal absorption lines.  They applied this technique
to the $\delta$~Scuti star FG~Vir and were able to assign $l$ values to the
eight strongest oscillation modes.  Two modes turned out to be radial,
allowing a match to theoretical models and hence precise determination of
mean density, luminosity, mass and distance (the latter of which agreed
very well with the Hipparcos parallax).  Work by \citet{BPP99} on FG~Vir
agrees quite well with that of \citeauthor{VKB98}.  \citeauthor{BPP99}
fitted all known 24 oscillation frequencies in FG~Vir to detailed models
using the mode identity of the 8 strongest modes.  Even when they included
convective overshoot and modified opacities, they were not able to match
all observed frequencies.  They identified a number of very good fits, but
none was perfect.  The reason is probably the treatment of rotation in the
models.
 
\subsection{Rotation in $\delta$~Scuti stars}

Rotation has a very strong effect on the observed frequency spectrum.
\citet{TBG2000} illustrated this nicely using model calculations on a
number of models representing the $\delta$~Scuti star $\theta$ Tucanae.
Based on 10 observed frequencies, they attempted to calculate the
rotationally split frequencies assuming uniform rotation.  They concluded
that although one is able to match the observed frequencies when
rotationally split frequencies are included in the models, one can actually
find several solutions if the rotational velocity is kept free in the
analysis.

It is important to note that rotation not only affects the frequencies, but
also the basic stellar properties.  This has been studied for stars in the
Praesepe cluster by \citet{MHH99}, who were able to correct for the
rotational effect on the measured temperature and luminosity.  They created
a complete seismic picture of all known $\delta$~Scuti stars in the
Praesepe star cluster, most of which were observed by the STEPHI network
\citep[e.g.,][]{HMB98}.  Although \citeauthor{MHH99} were able to correct
for rotation and, to some extent, limit the number of free parameters by a
kind of differential seismology between oscillating stars in the cluster,
they could not reach a satisfactory fit between the models and the observed
frequencies.
 
It is well known in the field of $\delta$~Scuti seismology that the stellar
rotation is probably the main problem.  Rotation affects the models (both
the stellar evolution and via frequency perturbations), implying that we
only calculate accurate theoretical frequencies if we are able to describe
the internal (differential!) rotation of the star.  As pointed out above,
rotation also affects the stellar parameters, so once again we can only
calculate accurate values for temperature and luminosity if we know the
rotation (and the shape) of the star.  Without this knowledge, we have
enough freedom to reach a reasonable fit to any combination of observed
frequencies.  However, even this freedom is not enough in most cases to
reach a perfect fit, which may be one of the most interesting conclusions
we can draw at present: the theoretical models do not describe
$\delta$~Scuti stars accurately enough to fit the observed oscillation
frequencies.
 
\section{Solar-like oscillations}

Observers have worked hard in the last decade on searching for solar-like
oscillations in other stars.  Reviews of those efforts have been given by
\citet{B+G94}, \citet{K+B95}, \citet{HJLaB96} and \citet{B+K98}.  The
problems which limit the usefulness of $\delta$~Scuti oscillation
frequencies do not apply to the solar-like case.  For one thing, most
solar-type stars are slow rotators.  Also, in the Sun we find that all
possible modes within a broad frequency range are excited.  The frequencies
form a fairly regular series which is well approximated by the so-called
asymptotic relation.  It is generally assumed that the same will apply to
other solar-like oscillations, which means that mode identification will
not be a problem.

The single most important problem with solar-like oscillations lies in
their extremely small amplitudes.  Detection therefore requires extremely
sophisticated techniques.  Such techniques are now available via
high-precision Doppler measurements, but a drawback is the difficulty are
arranging multi-site observations that are crucial to obtaining a decent
window function.
 
\subsection{Recent observational results}

We may define four different levels of detections in relation to solar-like
oscillations:
\begin{enumerate}\itemsep=0pt

\item No detection.
\item Detection of excess power.
\item Detection of average frequency separations ($\Delta\nu$ and
$\delta\nu_{02}$) by fitting to the asympotic relation.
\item Detection of individual frequencies (including mode identifications,
departures from the asympotic relation, rotational splitting, etc.).

\end{enumerate}

\begin{figure*}
\centering
\includegraphics[width=1.0\textwidth]{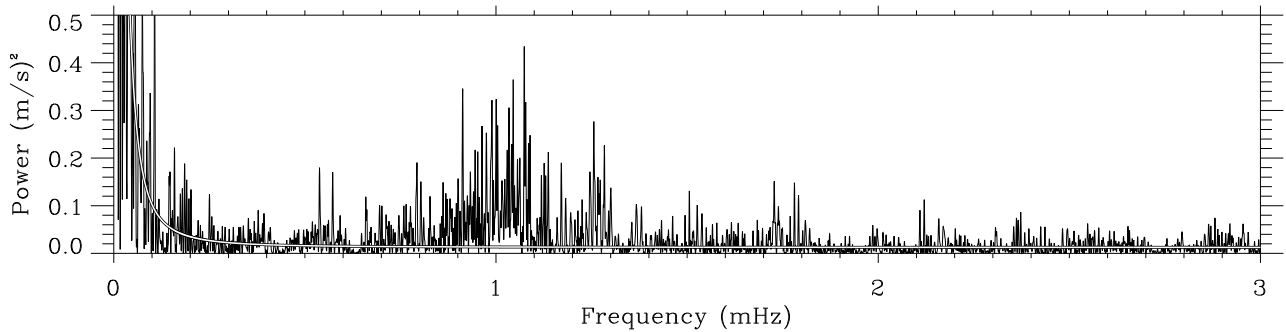}
\caption{\label{fig.bhyi} Power spectrum of AAT velocity measurements of
$\beta$ Hydri. There is a clear excess of power around 1 mHz which is a
striking signature of solar-like oscillations \citep[from][]{BBK2000}.}
\end{figure*}

Many unsuccessful attempts have been made during the last decade, all of
which must be placed at the first level.  Five years ago, a review by
\citet{K+B95} concluded there was little evidence for any published data to
be placed at any level higher than the first.  Such a conclusion cannot be
maintained today!  A number of observing campaigns have now produced
significant evidence for oscillations:
\begin{description}

\item[$\eta$~Boo] \citet{KBV95} detected excess power in this G0 subgiant
from measurements of Balmer-line equivalent widths.  The excess was at the
expected level, and these authors were able to extract frequency
separations and individual frequencies which agreed well with theoretical
models \citep{ChDBK95,G+D96}.  A more detailed discussion of theoretical
models for $\eta$~Boo can be found by Di Mauro \& Christensen-Dalsgaard
(these Proceedings).  We should note, however, that a search for velocity
oscillations in this star by \citet{BKK97} failed to detect a signal,
setting limits at a level below that expected on the basis of the
\citeauthor{KBV95} result.
 
\item[Procyon] Velocity data for Procyon (F5 subgiant) have recently
provided good evidence for oscillations \citep{BMM99,MSL99} with a clear
power excess around 1\,mHz and peak mode amplitudes of about
0.5\,m\,s$^{-1}$.  The large frequency separation ($\Delta\nu = 56\,\mu$Hz)
seems to agree with expectations, but the single-site window function
complicates the interpretation.  These results indicate that the power
excess seen earlier by \citet{BGN91} --- which implied similar peak
amplitudes --- may have been real, although the inferred $\Delta\nu$ of
71\,$\mu$Hz does not agree with the more recent result.

\item[$\zeta$ Herculis] Martic et al.\ showed evidence at this conference
for p-mode power in this G0 subgiant.  A comb-response analysis apparently
shows a large separation of 43.1\,$\mu$Hz.  Based on their echelle diagram,
one may even extract a value for the small separation of $\delta\nu_{02} =
3\,\mu$Hz.  Again, the single-site window funciton is problematic.

\item[$\alpha$~Cen~A] \citet{KBF99} measured Balmer-line equivalent widths
in $\alpha$~Cen~A with two telescopes over six nights and set an upper
limit on oscillation amplitudes of only 1.4 times solar, with tentative
evidence for p-mode structure.  Photometry from the WIRE satellite by Schou
and Buzasi (these Proceedings) appears to confirm oscillations at
approximately this level.

\item[$\beta$~Hyi] \citet{BBK2000} have made what seems to be the best
example of a detection of solar-like oscillations in another star.  Their
power spectrum of the G2 subgiant $\beta$ Hydri can be seen in
Fig.~\ref{fig.bhyi}.  The star was observed in velocity over five nights
with the UCLES echelle spectrograph on the 3.9-m Anglo-Australian Telescope
(AAT), using an iodine cell as the wavelength reference.  The power excess
is at the expected level and a fit to the asymptotic relation results in a
large separation of 56.2\,$\mu$Hz.  The oscillation frequencies seem to
depart significantly from the asymptotic relation, but again there are big
problems from the single-site window function.

\end{description}

Given that oscillations are now being reliably detected, it seems clear to
us that the usefulness of single-site observations is severely limited.
One may even say that further observations with single telescopes would be
a waste of telescope time.  The time has come to organise campaigns with
two or more telescopes, and perhaps to think about a dedicated network.

\subsection{Amplitudes of solar-like oscillations}

Solar-like oscillations are excited by convection and the expected
amplitudes have been estimated using theoretical models.  Based on models
by \citet{ChD+F83}, \citet{K+B95} suggested that amplitudes in velocity
should scale as $L/M$, where $L$ and $M$ are the stellar luminosity and
mass in solar units.  More recent calculations by \citet{HBChD99} confirm
this scaling relation, at least for stars with near-solar effective
temperatures.  However, it was pointed out by \citet{K+B95} that the $L/M$
relation predicted amplitudes for some F-type stars, namely Procyon and
several members of the cluster M67, that were greater than observational
upper limits.  Despite this, the $L/M$ relation has been quite widely
adopted.

In order to resolve the problem for the hotter stars, we suggest a revised
scaling relation.  Noting that $L/M$ is equal to $T_{\rm eff}^4/g$ (with
all quantites in solar units), we propose a modified relation in which
velocity amplitudes scale as $1/g$.  Given the growing number of credible
detections, we are now able to check these relations.  Note that, although
a single-site window function makes it difficult to extract frequencies,
the estimates of oscillation amplitudes are not so badly affected.

\begin{figure}
\centering
\includegraphics[width=0.45\textwidth]{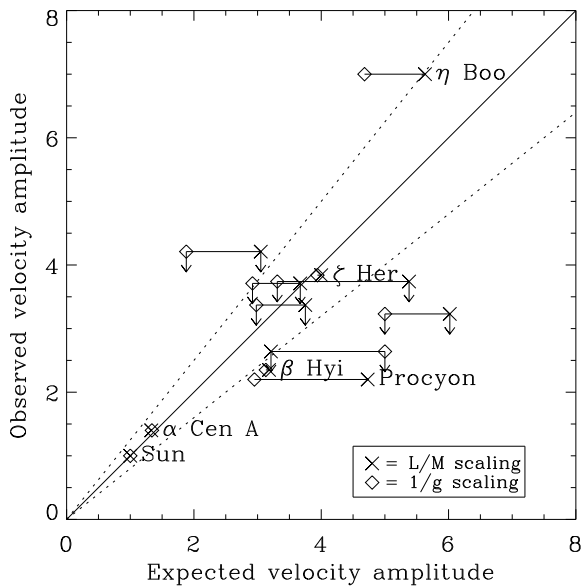}
\caption[]{\label{fig.amp} Observed amplitudes for solar-like oscillations
compared with predictions based on the $L/M$ and $1/g$ scaling relations.
All amplitudes are relative to solar.  The upper limits are for stars in
the cluster M67.  The diagonal line shows equaliy, with the dotted lines
showing $\pm$25\%.}
\end{figure}

Figure~\ref{fig.amp} shows the observed versus expected velocity amplitudes
for the stars mentioned above, using both scaling relations.  The upper
limits in the figure are from photometric observations of stars in the open
cluster M67 \citep{GBK93,K+B95}.  Those results, as well as observations in
equivalent width ($\eta$~Boo and $\alpha$~Cen~A), have been converted to
velocity amplitudes using the relations given by \citet{K+B95}.

For the G-type stars ($\eta$~Boo, $\alpha$~Cen~A, $\beta$~Hyi, $\zeta$~Her)
there is, of course, little difference between the two scaling relations.
The agreement with observations is generally good.  More data are clearly
needed, but it is comforting to see that these stars do indeed appear to be
oscillating at the expected levels.  Turning to F-type stars, we see that
the $1/g$ relation gives a much better prediction for the amplitude of
Procyon, and also relaxes the problematic upper limits for the stars in
M67.  Again, more data are needed, but in the mean time we suggest using
$1/g$ rather than $L/M$ to predict oscillations amplitudes.

\section{Space missions}

The recent progress in ground-based observations, as well as results from
the 52\,mm star tracker on the otherwise-failed WIRE satellite \citep[][and
these Proceedings]{BCL2000}, illustrate the potential of upcoming space
missions:
\begin{description}\itemsep=0pt  \vspace*{-2ex}

\item[MOST]  A Canadian project \citep{MKW2000}.

\item[MONS] A Danish-led project with contributions from Australia and
other countries \citep*{KBChD2000}.

\item[COROT] A French/European project \citep{Bag98}.

\item[Eddington] A proposed ESA Flexi-mission, which has been selected as a
reserve mission.

\end{description}
 
\begin{figure}
\centering \includegraphics[width=0.47\textwidth]{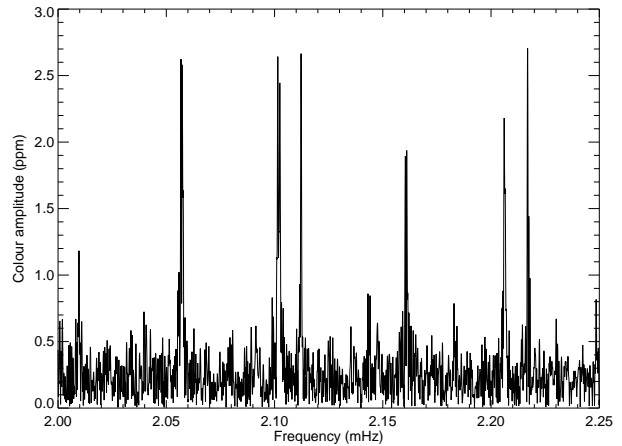}
\caption{\label{fig.acen} Simulated power spectrum for $\alpha$~Cen~A, as
expected from the MONS space mission.  }
\end{figure}

The future of asteroseismology is in space.  One can perform wide-band
photometry with high accuracy because of the absence of scintillation
(fluctuations in the stellar light caused by the Earth's atmosphere).  Even
a small space-based telescope will do much better than even the largest
ground-based telescopes.  The other major improvement from moving to space
is an excellent window function that is unaffected by weather.  In space
one can reach duty cycles of 80--90\%, which will result in very low
side-lobes.  As an example, we show in Fig.~\ref{fig.acen} a simulated data
set for $\alpha$~Cen~A as observed by MONS\@. 

\section{Conclusions}

In this review we have described the current status of asteroseismology for
A, F and G subgiants and main sequence stars.  The main conclusions are as
follows:
\begin{itemize}

\item Asteroseismology is far from helioseismology in terms of techniques
and results.

\item There are plenty of challenges for theoreticians, such as mixing,
fluid motions, turbulent convection and deviations from spherical symmetry.
The most important area to focus on is rotation.

\item Asteroseismology is observationally driven.  Ground-based velocity
observations are now achieving believable detections, and it is time for
multi-site campaigns.  A large ground-based network is very desirable.

\item A number of exciting space missions are in various stages of planning
and construction (MOST, COROT, MONS and Eddington).  If these succeed, it
seems likely that asteroseismology will enter a golden age.
 
\end{itemize}

\section*{Acknowledgements}
 
This work was supported by the Danish National Research Foundation through
its establishment of the Theoretical Astrophysics Center and by the
Australian Research Council.

\end{document}